%% file: Async_Error_Floor_CM.tex
\documentclass[10pt,conference]{IEEEtran}

\IEEEoverridecommandlockouts

\usepackage{cite}
\usepackage{amsthm}
\usepackage{amsmath,amssymb,amsfonts}
\usepackage{algorithmic}
\usepackage{graphicx}
\usepackage{textcomp}
\usepackage{xcolor}
\usepackage{bm}
\usepackage{todonotes}
\usepackage{tabu}
\usepackage{balance}
\def\BibTeX{{\rm B\kern-.05em{\sc i\kern-.025em b}\kern-.08em
    T\kern-.1667em\lower.7ex\hbox{E}\kern-.125emX}}

\usepackage{acro}
\usepackage{mathtools}
\mathtoolsset{showonlyrefs}
\input{acronyms2.tex}

\begin{document}
\title{Error Floor Analysis\\of Irregular Repetition ALOHA}

\author{\IEEEauthorblockN{Federico Clazzer\IEEEauthorrefmark{1} and Alexandre Graell i Amat\IEEEauthorrefmark{2}}
\IEEEauthorblockA{\small\IEEEauthorrefmark{1}Institute of Communications and Navigation, German Aerospace Center (DLR), Wessling, Germany\\
\small\IEEEauthorrefmark{2}Department of Electrical Engineering, Chalmers University of Technology, Gothenburg, Sweden
\vspace{-.57cm}}
\thanks{A.~Graell i Amat was supported by the Swedish Research Council under grant 2020-03687.}}

\maketitle

\begin{abstract}
With the rapid expansion of the Internet of Things, the  efficient sharing of the wireless medium by a large amount of simple transmitters is becoming essential. Scheduling-based solutions are inefficient for this setting, where small data units are broadcast sporadically by terminals that most of the time are idle. Modern random access has embraced the challenge and provides suitable slot-synchronous and asynchronous multiple access solutions based on replicating the packets and exploiting \ac{SIC} at the receiver. In this work, we focus on asynchronous modern random access. Specifically, we derive an analytical approximation of the performance  of \ac{IRA} in the so-called error floor region. Numerical results show the tightness of the derived approximation under various scenarios.
\end{abstract}


\input{notation.tex}

\section{Introduction}
\acresetall
Under the label of \ac{IoT} and \ac{M2M}, a number of services revolutionizing vast industry sectors, from automotive to logistics, from health care to farming, have emerged in the last years \cite{Aceto2019, Philip2021}. Many of these  services  are characterized by the sporadic transmission of data units containing few information bits. This apparently small twist in the communication problem poses a number of intriguing challenges. At the physical layer, efficient transmission detection, channel estimation, and error correction for short packets are key. Further, an extensive transmitter population needs to share the common wireless medium as efficiently as possible. Very low duty cycle, sporadic access, and small amount of information are ineffectively handled by classical scheduling-based \ac{MAC} approaches. \Ac{RA} inherently provides the sought flexibility with the shortcoming of low efficiency \cite{Abramson1970}.

\emph{Modern random access} (see \cite{Casini2007,Liva2011,Paolini2014,Sandgren2017}) has shown how the use of packet replication coupled with \ac{SIC} is able to drastically boost performance. Tools borrowed from the design of \ac{LDPC} codes over the erasure channel can be exploited to maximize the system throughput and overcome the low efficiency burden of classical schemes. A  random coding bound for the massive random access setting was derived in \cite{Polyanskiy2017}. In the quest to close the gap emerged with respect to modern random access solutions, the research community has also investigated  compressed-sensing inspired schemes \cite{Amalladinne2020, Fengler2020, Truhcachev2020}.

Slot-synchronous modern \ac{RA} have been extensively studied both in the asymptotic regime\cite{Liva2011,Paolini2014}, i.e., when the maximum delay among physical layer packet copies grows very large, and in the finite length regime \cite{Sandgren2017,Graell2018_FL}. Rendering the entire transmitter population synchronized to the slot boundaries entails a certain level of complexity and requires the senders to be able to receive a beacon signal. Such requirements might be undesirable for low-complexity and low-cost \ac{IoT} applications where terminals can be equipped with  transmitter-only hardware and are battery-powered. Asynchronous \ac{RA} dates back to the original idea of Abramson's  ALOHA protocol \cite{Abramson1970} and has also been investigated in the context of modern \ac{RA}. The use of packet copies\textemdash referred to as replicas in the following\textemdash spaced with a randomized delay and the use of \ac{SIC} at the receiver has proven to be particularly beneficial \cite{Kissling2011a,deGaudenzi2014_ACRDA}. Packet combining can be used to supplement \ac{SIC} and further improve performance \cite{Clazzer18:ECRA}. The extension to a variable number of replicas per user, dubbed \ac{IRA}, was introduced in \cite{Akyldiz2020}.

The asymptotic setting of asynchronous modern \ac{RA} was studied in \cite{Akyldiz2020} under the simplified destructive collision channel model. An error floor analysis  was provided in \cite{Clazzer18:ECRA} for the simplest collision pattern neutralizing \ac{SIC} with two users and two replicas per user.

In this work, we derive an analytical approximation  to the performance of \ac{IRA} in the error floor region, i.e., for low-to-medium channel loads. Compared to the analysis in \cite{Clazzer18:ECRA}, the derived approximation is very general and encompasses \ac{IRA} with different number of replicas per user as well as collision patterns with more than two packets. For the particular case of two replicas per user, it closes the gap to the simulated performance of the prediction in \cite{Clazzer18:ECRA}. The derived analytical approximation yields a very accurate prediction of \ac{IRA} in the error floor region, and can be used for the optimization of the degree distribution according to which users select the replication factor to achieve a target packet loss rate.

\section{System model}
\label{sec:sys_ov}

\begin{figure}[!t]
	\centering
	\includegraphics[width=\columnwidth]{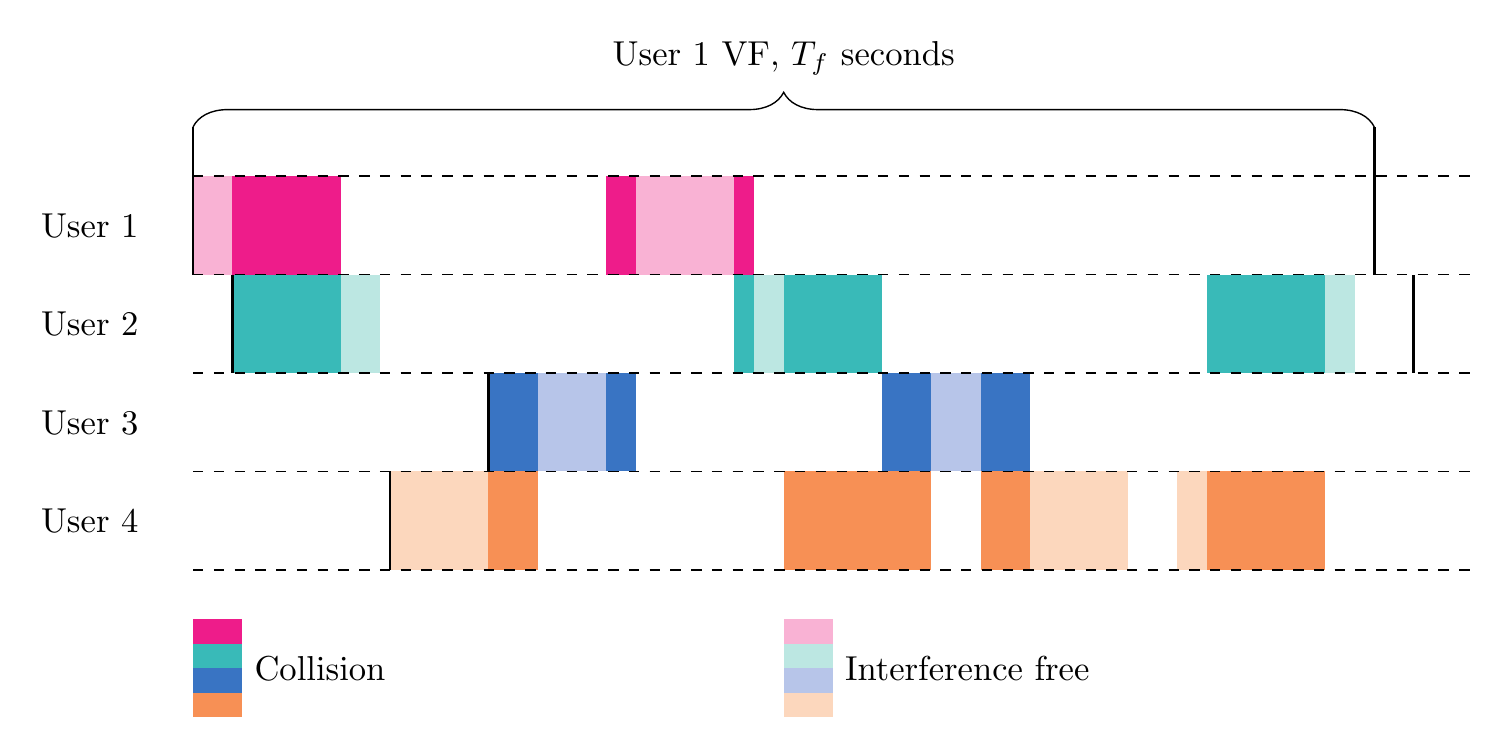}
	\vspace{-5ex}
	\caption{Four users transmit  their replicas according to the \ac{IRA} protocol. User $1$ and $3$ send two replicas,  User $2$  three replicas, and User $4$ four replicas.}
	\label{fig:tx_sig}
	\vspace{-3ex}
\end{figure}

We consider an infinitely large user population generating traffic according to a Poisson process of intensity $\lo$, called \emph{logical channel load} or simply \emph{load} in the following. $\lo$ is measured in packet arrivals per packet duration $\pkLen$. Further, we assume the  \ac{IRA} asynchronous RA protocol  \cite{Kissling2011a,Akyldiz2020}. According to \ac{IRA},  when a packet is generated, the user samples a  degree distribution $\ud$ to compute the repetition degree $\dg^{(\us)}$. The polynomial representation of the user degree distribution is of the form \begin{align*}
\ud(w)=\sum_{\dg=2}^{\dgM} \ud_\dg w^\dg\,,
\end{align*}
where $\ud_\dg$ is the probability that $\dg$ replicas are transmitted, $\dgM$ is the maximum number of packet copies sent, ${\sum_{\dg=2}^{\dgM} \ud_\dg=1}$, and  $\dgm=\sum_{\dg=2}^{\dgM} \dg \ud_\dg$ is the average number of replicas per data unit transmitted.\footnote{The logical channel load $\lo$ is the measure of innovative packets injected by the user population, equivalent to the channel load of classical \ac{RA} schemes like ALOHA and \ac{SA}, and thus does not take into account the user degree distribution. The \emph{physical channel load} instead measures the average number of packets per packet duration effectively transmitted over the channel and thus is computed as $\lo_p=\lo \dgm$.} While the first replica is transmitted immediately upon generation of the data unit of user $\us$, the remaining $\left(\dg^{(\us)}-1\right)$ replicas are sent within a \ac{VF} of duration $\fraLen$, with transmission times chosen so as to avoid self-interference. Since packets are generated at random with exponential inter-arrival time, the \acp{VF} are asynchronous among users. The start time of each replica of a user can be stored in a dedicated portion of the packet header or  uniquely determined from the data content via a pseudo-random algorithm known at the receiver \cite{Clazzer18:ECRA}. An example  of a received signal with four users transmitting two, three, two, and four replicas, respectively, is depicted in Fig.~\ref{fig:tx_sig}.

\begin{sloppypar}
To withstand the effect of the Gaussian channel and interference, replicas are protected by a channel code $\Code$
with Gaussian codebook. Its code rate is $\rate=\numBit/\numSym$, where $\numSym$ is the number of  packet symbols after channel encoding and modulation. Assuming an ideal estimate of the sampling epoch, the frequency offset, and the phase offset, the discrete-time baseband  received signal for the $\replica$-th replica of the $\us$-th user, ${\rxVec^{(\us,\replica)} = (\rx_0^{(\us,\replica)}, ..., \rx_{\numSym-1}^{(\us,\replica)})}$, is given by
\begin{equation}
\rxVec^{(\us,\replica)} = \symVec^{(\us)} + \intVec^{(\us,\replica)} + \noiseVec\,.
\end{equation}
Here $\symVec^{(\us)} = \left(\symVal_{0}^{(\us)}, \dots, \symVal_{\numSym-1}^{(\us)}\right)$ is the sequence of transmitted symbols, $\intVec^{(\us,\replica)}$  the interference contribution over the user-$\us$ replica-$\replica$ signal and $\noiseVec=\left(\noise_0, \dots, \noise_{\numSym-1}\right)$  a noise vector sampled from a complex discrete white Gaussian process with ${\noise_i \sim \mathcal{CN}(0,2\noiseVar)}$.
\end{sloppypar}

For user $\us$, replica $\replica$, and symbol $i$, we define ${\Pw_{i}^{(\us)} \triangleq \mathbb{E}\left[|\symVal_{i}^{(\us)}|^2 \right]}$, the received signal power, $\Ns=2\noiseVar$ the noise power, and ${\intPw_{i}^{(\us, \replica)} \triangleq \mathbb{E}\left[ |\intVal_{i}^{(\us,\replica)}|^2 \right]}$ the aggregate interference power. Throughout the paper, we assume that all users are received with the same power, i.e. perfect power control is adopted. Thus,  $\Pw_{i}^{(\us)} = \Pw$ and $\intPw_{i}^{(\us, \replica)} = \intNum_i^{(\us, \replica)} \Pw$, where $\intNum_i^{(\us, \replica)}\in \mathbb{N}$ denotes the number of active interferers over the $i$-th symbol. The aggregate interference is a discrete Gaussian process, with $\intVal_{i} \sim \mathcal{CN}\left(0,\intNum_i^{(\us, \replica)} \Pw\right)$. Hence, the instantaneous \ac{SINR} is
\begin{equation}
\sinr_{i}^{(\us, \replica)} = \frac{\Pw}{\Ns + \intNum_i^{(\us, \replica)} \Pw}\,.
\end{equation}

\subsection{Modeling of the decoding process}
\label{sec:int_model}
Instead of adopting the well-known destructive collision channel model that is particularly pessimistic in the case of asynchronous access in the presence of error correction, we resort to the \emph{block interference channel model} \cite{McEliece1984}. Consider again the $\replica$-th replica of  user $\us$. To ease the notation, we drop the superscript ${(\us, \replica)}$. We interpret the $\numSym$ symbols as independent Gaussian channels and, leveraging the Gaussian assumption of both the transmitted signals and the noise, we compute the instantaneous mutual information over the $i$-th symbol, $\mutInf(\sinr_{i})$, as
\begin{equation}
\mutInf(\sinr_{i}) = \log_2( 1 + \sinr_{i} )\,.
\end{equation}
The average mutual information $\avMutInf$ over the $\numSym$ symbols is\footnote{Similar to \cite{Thomas_2000}, the idea is to take into account the mutual information carried by each replica symbol, and then compute the average over the entire replica.}
\begin{equation}
\label{eq:mi_dec}
\avMutInf = \frac{1}{\numSym} \sum_{i=0}^{\numSym-1} \mutInf(\sinr_{i}) = \frac{1}{\numSym} \sum_{i=0}^{\numSym-1} \log_2( 1 + \sinr_{i} )\,.
\end{equation}
By comparing the average mutual information over the packet with the rate $\rate$, we resort to the decoding condition $D = \mathbb{I}\left\{\rate \leq \avMutInf\right\}$, where $\mathbb{I}\{X\}$ denotes the indicator function. Hence,   $D=1$  if decoding is  successful and $D=0$ otherwise.

The destructive collision channel model is a special case of the block interference channel model, where $\rate$ is chosen such that only collision-free packets can be successfully decoded, i.e., $\rate = \log_2\left(1+\frac{\Pw}{\Ns}\right)$. In asynchronous \ac{RA} solutions, to be able to counteract a certain level of interference, a rate $\rate < \log_2\left(1+\frac{\Pw}{\Ns}\right)$ is normally chosen. Additionally, this model allows to take into account features like channel coding, multi-packet reception, and capture effect \cite{Ghez1988, Zorzi1994}.

In \ac{IRA}, when the decoding of a replica is successful, its contribution together with the one of all its copies is removed from the received signal via ideal interference cancellation.\footnote{Ideal packet detection is assumed in the following. Through a careful selection of the preamble, the detection algorithm and the detection threshold, misdetections can be minimized.}

\section{Analytical approximation of the packet loss rate in the error floor region}

In this section, we derive an analytical approximation of the \ac{PLR} of \ac{IRA} that is tight for low-to-medium channel load conditions. Packet losses are caused by particular interference patterns that \ac{SIC} is not able to resolve. In the slot-synchronous case, these patterns are analogous to the stopping sets of \ac{LDPC} codes \cite{Di2002}, and can be analyzed exploiting tools from coding theory and graph theory. In the asynchronous setting, a graph representation is not straightforward, since no discrete objects as slots are present anymore. Building on the error floor approximation in \cite{Iva2015, Sandgren2017}, we derive a tighter \ac{PLR} approximation than the one in \cite{Clazzer18:ECRA} by considering a larger subset of \acp{UCP}, beyond the two-user case considered in \cite{Clazzer18:ECRA}. Enlarging the considered subset not only improves the approximation for regular user degree distributions, but also enables the analysis of irregular user degree distributions for the first time. The following definitions will be useful for the analysis.
\begin{definition}[\textbf{Collision cluster $\CollClus$}]
Consider a subset $\CollClus$ of users. Assume that packets of all users in the complement set of $\CollClus$, denoted by $\CollClus^{c}$, have been successfully decoded. The subset $\CollClus$ is referred to as \emph{collision cluster} if no packet replicas for the users in $\CollClus$ are collision-free.
\end{definition}
Under the assumption of a collision channel, none of the users in the collision cluster can be successfully decoded. Conversely, when a channel code $\Code$ protects each transmitted packet, a certain level of interference can be tolerated yet allowing correct decoding and thus the collision cluster might be resolvable.
\begin{definition}[\textbf{$\Code$-unresolvable collision pattern}]
\acreset{UCP}
Assuming that each packet is encoded by a code $\Code$, a \emph{\ac{UCP}} $\UCP$ is a collision cluster where no user in the set can be successfully decoded.
\end{definition}
A \ac{UCP} is also a collision cluster, but not vice versa. For low-to-medium channel loads, decoding failures are caused by  \ac{UCP}s  involving few users and correspond to the minimal stopping sets in the slot-synchronous scenario \cite{Iva2015, Ivanonv17:alltoall}.
\begin{definition}[\textbf{Dominant $\Code$-unresolvable collision pattern}]
A dominant \ac{UCP} does not contain a nonempty \ac{UCP} of smaller size.
\end{definition}
In order to evaluate the probability of \ac{UCP}s, we extend the definition of vulnerable period \cite{Kleinrock1976_book} as follows:
\begin{definition}[\textbf{$\Code$-vulnerable period for $|\UCP|=2$}]
Consider the transmission of a packet protected by a code $\Code$ between time $\RStart$ and $\RStart + \pkLen$. The packet's \emph{$\Code$-vulnerable period} is the interval of time $\left[\RStart - \VL, \RStart + \VR\right]$ in which the presence of a single interferer leads to a decoding failure. Hence, the vulnerable period duration $\Vpd$ is defined as $\Vpd = \VL + \VR$.
\end{definition}
Under the collision channel model, $\VL= \VR= \pkLen$, so ${\Vpd = 2 \pkLen}$, as known from the literature \cite{Kleinrock1976_book}. Instead, when packets are protected by a code $\Code$ characterized by a rate $\rate$ [b/symb], some interference can be sustained before decoding may fail, effectively reducing the vulnerable period to
\begin{equation}
\label{eq:vp}
\Vpd = 2\ifr \pkLen, \qquad 0 \leq \ifr \leq 1\,.
\end{equation}
Leveraging on the Gaussian assumption of both the signals and noise, $\ifr$ is the unique solution of the equation
\begin{equation}
\label{eq:phi0}
\ifr \log_2\left(1 + \frac{\Pw}{\Ns}\right)+( 1 - \ifr ) \log_2 \left(1 + \frac{\Pw}{\Ns + \Pw}\right) = \rate
\end{equation}
when it exists. Hence,
\begin{equation}
\label{eq:phi}
\ifr = \max\left(0, \frac{\rate-\log_2 \left(1 + \frac{\Pw}{\Ns + \Pw}\right)}{\log_2\left(1 + \frac{\Pw}{\Ns}\right) -\log_2 \left(1 + \frac{\Pw}{\Ns + \Pw}\right)}\right)\,,
\end{equation}
with $0 \leq \ifr \leq 1$.\footnote{The present analysis can be easily extended to the block fading channel. Conditioning on the received power for the replica under investigation and the power of the interfering data unit, \eqref{eq:phi0} still holds. By removing the conditioning over the two power levels, one can retrieve the density of $\ifr$ and hence its mean. This value can then be used for the computation of the vulnerable period required in the packet loss rate approximation.}

An \ac{UCP} is characterized by a number of replica-collision sets where the users following the user profile $\up^\UCP$ transmitted their replicas. Both the replica-collision set and $\up^\UCP$ are defined as follows.
\begin{definition}[\textbf{Replica-collision set}]
Consider the \ac{UCP}~$\UCP$. A replica-collision set within \ac{UCP} collects all packets that cause reciprocal interference, i.e., no packet within a replica-collision set is interference-free and its collision is with one or more data units of the replica-collision set.
\end{definition}

We denote by $\spf^\UCP$ the number of replica-collision sets in   \ac{UCP}~$\UCP$.

\begin{definition}[\textbf{User profile $\up^\UCP$ for \ac{UCP} $\UCP$}]
Consider the \ac{UCP}~$\UCP$. The user profile $\up^\UCP$ is defined as the vector $\up^\UCP=\left[\upf_1^\UCP, \cdots, \upf_{\dgM}^\UCP\right]$, where $\upf_{l}^\UCP$ is the number of users in $\UCP$ transmitting $l$ replicas. We denote by $\upf^\UCP$ the $L^1$-norm of $\up^\UCP$, i.e., $\upf^\UCP=\parallel\up^\UCP \parallel_1=|\UCP|$.
\end{definition}

\begin{figure}[!t]
	\centering
	\includegraphics[width=\columnwidth]{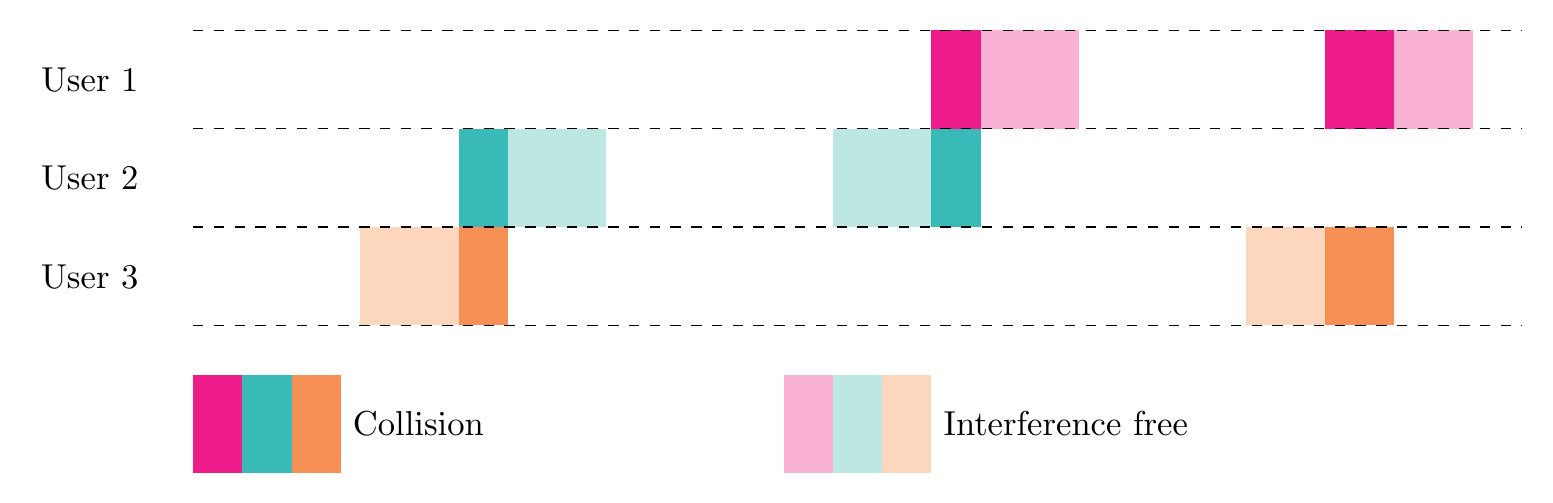}
	\vspace{-4ex}
	\caption{Example of a \ac{UCP} characterized by $\spf^\UCP=3$ replica-collision sets. From left to right,  the first set involves user $2$ and $3$, the second user $1$ and $2$ and the last one user $1$ and $3$. The number of involved users in the \ac{UCP} is $\upf^\UCP=3$ and the user profile follows $\up^\UCP=[0,3,0,0]$, i.e. all three users transmitted $2$ replicas.}
	\label{fig:S3}
	\vspace{-3ex}
\end{figure}

An example of a \ac{UCP} is depicted in Fig.~\ref{fig:S3}. For this example, the \ac{UCP} contains  $\spf^\UCP=3$   replica-collision sets, the number of involved users is $\upf^\UCP=3$, and the user profile is $\up^\UCP=[0,3,0,0]$. This set will be denoted by $\UCP_3$ in the following (see~Table~\ref{tab:UCP}).

\subsection{Packet loss rate approximation}

We are now ready to derive an approximation to the \ac{PLR} for \ac{IRA}. Consider a generic user~$\us$. Denote by $\AllUCP$ the set of all \ac{UCP}s, with $\DUCP \subset \AllUCP$ the set of dominant \ac{UCP}s, and by $\s=\fraLen/\pkLen$ the \ac{VF} time span measured in packet durations. Similar to  \cite{Iva2015,Sandgren2017,Clazzer18:ECRA}, the \ac{PLR}~$\plr$ can be approximated as
\begin{align}
	\plr &= \Pr\left(\bigcup_{\UCP \in \AllUCP} \us \in \UCP \right)\stackrel{(a)}{\leq} \sum_{\UCP \in \AllUCP}\Pr\left(\us \in \UCP \right) \nonumber\\ &\stackrel{(b)}{\approx}\sum_{\UCP \in \DUCP}\Pr\left(\us \in \UCP \right) \nonumber \\
	&\stackrel{(c)}{=}\sum_{\tus=2}^{\infty}\sum_{\UCP \in \DUCP} \ps \Pr(M=\tus) \nonumber\\
	& = \sum_{\tus=2}^{\infty}\sum_{\UCP \in \DUCP} \ps \frac{e^{-\s \lo}(\s \lo)^\tus}{\tus!}\,,
\label{eq:PLR}
\end{align}
where $(a)$ is the well-known union bound, $(b)$ stems from approximating the \ac{PLR} by considering only the dominant $\Code$-unresolvable collision patterns, and $(c)$ follows by conditioning on having $m$ users transmitting in a \ac{VF} time span. Note that the truncation of the union bound is by construction an approximation. Further, focusing on the \ac{UCP} contributing the most to the \ac{PLR} in the low-to-moderate channel load region is a natural choice since enumerating all possible \acp{UCP} is infeasible. Additionally, for low channel load values it is clear that \acp{UCP} with a small number of users involved are more likely to occur, hence $\DUCP$ will comprise only the dominant \acp{UCP} with a small number of transmitters involved.

We further define $\nVp=\lfloor \fraLen/\Vpd \rfloor$, the number of disjoint vulnerable periods within one \ac{VF}. Exploiting the parallelism between \acp{UCP} for the asynchronous case and stopping sets for the slot-synchronous case, following \cite{Sandgren2017}, $\ps$ can be written as
\begin{align}
\label{eq:pru}
\ps=\frac{\as \,\bs \,\cs}{\ds}\cdot\frac{\upf^\UCP}{m}\,,
\end{align}
where $\as$ is the number of ways to select $\upf^\UCP$ users with degree profile $\profileS$ from a set of $\tus$ users with degree distribution $\Lambda(w)$, $\bs$ is the number of ways to select the vulnerable periods of $\UCP$ such that $u \in \UCP$, $\cs$ is the number of graph-isomorphisms of $\UCP$,
and $\ds$ is the total number of ways in which $\upf^\UCP$ users (including $u$) with degree profile $\profileS$ can connect edges to the $\nVp$ (disjoint) vulnerable periods in their \acp{VF}. These terms can be easily obtained via combinatorial arguments as \cite{Sandgren2017}
\begin{align}
\as &= \binom{\tus}{\upf^\UCP}  \upf^\UCP! \prod_{l=2}^{\dgM}\frac{\left( \Lambda_l\right)^{\upf_l^\UCP}} {\upf_l^\UCP !}\,, \label{eq:as}\\
\bs & \approx \binom{\nVp-1}{\spf^\UCP-1}\,,\label{eq:bs}\\
\ds & \approx \frac{1}{\nVp}\prod_{l=2}^{\dgM} \left(\nVp\binom{\nVp-1}{l-1}\right) ^{\upf_l^\UCP}\label{eq:ds}\,.
\end{align}

Using \eqref{eq:as}, \eqref{eq:bs} and \eqref{eq:ds} in \eqref{eq:pru} and substituting \eqref{eq:pru} in \eqref{eq:PLR}, we get the sought \ac{PLR} approximation. Note that when computing $\bs$ and $\ds$, we leverage the definition of disjoint vulnerable periods within a \ac{VF} by $\nVp=\lfloor \fraLen/\Vpd \rfloor$. These are the equivalent of time slots in slot-synchronous \ac{RA}. The duration is computed according to \eqref{eq:phi}, which assumes the collision among only two packets and therefore is exact only for \ac{UCP}s with $\upf^\UCP=2$. For \ac{UCP}s involving more than two users there is a nonzero probability that collisions among more than two replicas appear. Nonetheless, we conjecture that considering only the case of collisions among two packets for the computation of $\nVp$ provides a good approximation, and we provide quantitative proof via the numerical results presented in the next section.

For a regular user degree distribution with degree $\dg$, $\as$ and $\ds$ simplify to
\begin{align}
\as &= \binom{\tus}{\upf^\UCP} \label{eq:asc}\\
\ds &\approx \frac{1}{\nVp}\left(\nVp\binom{\nVp-1}{\dg-1}\right)^{\upf^\UCP}\label{eq:dsc}\,,
\end{align}
leading to the approximation of the \ac{PLR}
\begin{align}
\plr \approx \sum_{\tus=2}^{\infty}\sum_{\UCP \in \DUCP}  \frac{\upf^\UCP \binom{\tus}{\upf^\UCP}   \binom{\nVp-1}{\spf^\UCP-1} \cs}{\frac{m}{\nVp} \left(\nVp\binom{\nVp-1}{\dg-1}\right)^{\upf^\UCP}} \frac{e^{-\s \lo}(\s \lo)^\tus}{\tus!}\,.
\label{eq:ubCRDSA}
\end{align}

For the particular case where we consider only the single unresolvable collision pattern consisting of two users, $\bar{\UCP}$, it follows $\upf^{\bar{\UCP}}=2$, $\spf^{\bar{\UCP}}=\dg$, and $c(\bar{\UCP})=1$, and \eqref{eq:ubCRDSA} further simplifies to
\begin{align}
\label{eq:ubCRDSAb}
\plr \approx \sum_{\tus=2}^{\infty}  \frac{\binom{\tus}{2}}{\nVp\binom{\nVp-1}{\dg-1}} \cdot\frac{2}{\tus} \cdot\frac{e^{-\s \lo}(\s \lo)^\tus}{\tus!}\,,
\end{align}
which is the expression in \cite[eq.~(18)]{Clazzer18:ECRA}.

\section{Numerical Results}

\begin{table}[!t]
\caption{Parameters of Dominant \ac{UCP}s with $\spf^\UCP \leq 4$}
\vspace{-2ex}
\centering
{\tabulinesep=1mm
\begin{tabu}{c|c|c|c|c}
\hline \hline
$\DUCP$ & $\up^\UCP$ & $\upf^\UCP$ & $\spf^\UCP$ & $\cs$\\
\hline \hline
$\UCP_1$ & $[0,2,0,0]$ & $2$ & $2$ & $1$\\
$\UCP_2$ & $[0,0,2,0]$ & $2$ & $3$ & $1$\\
$\UCP_3$ & $[0,3,0,0]$ & $3$ & $3$ & $6$\\
$\UCP_4$ & $[0,2,1,0]$ & $3$ & $3$ & $6$\\
$\UCP_5$ & $[0,0,0,2]$ & $2$ & $4$ & $1$\\
$\UCP_6$ & $[0,2,0,1]$ & $3$ & $4$ & $6$\\
$\UCP_7$ & $[0,1,2,0]$ & $3$ & $4$ & $12$\\
$\UCP_8$ & $[0,1,1,1]$ & $3$ & $4$ & $12$\\
$\UCP_9$ & $[0,0,3,0]$ & $3$ & $4$ & $24$\\
$\UCP_{10}$ & $[0,0,2,1]$ & $3$ & $4$ & $12$\\
$\UCP_{11}$ & $[0,3,0,1]$ & $4$ & $4$ & $24$\\
$\UCP_{12}$ & $[0,4,0,0]$ & $4$ & $4$ & $72$\\
\hline
\end{tabu}}
\label{tab:UCP}
\vspace{-4ex}
\end{table}

In this section, numerical results for the \ac{PLR} approximation derived in \eqref{eq:PLR} are provided. The set of twelve dominant \ac{UCP}s $\DUCP$ taken into account for the approximation is listed in Table~\ref{tab:UCP}.\footnote{Depending on the user degree distribution, $\DUCP$ may include only the subset of \ac{UCP}s that are viable. As an example, for $\ud=x^2$, only the subset $\{\UCP_1,\UCP_3,\UCP_{12}\}$ contributes to the \ac{PLR} approximation.} We consider perfect power control, i.e., all users are received with equal power, and a \ac{SNR} of $\SNR=6$ dB. As per \cite{Clazzer18:ECRA}, the receiver operates on a sliding window of duration $3\, \fraLen$ and is moved by $0.1\, \fraLen$ forward when no further packets can be successfully recovered in the current decoding window. We plot the \ac{PLR} for  two regular user degree distributions, $\ud=x^2$ and $\ud=x^3$, named \mbox{IRA-2} and \mbox{IRA-3} in the presented results, respectively, and for two irregular user degree distributions, namely $\ud_1=0.263x^2+0.344x^3+0.393x^5$ and $\ud_2=0.51x^2+0.49x^4$.

\begin{sloppypar}
In Fig.~\ref{fig:n200_R=1.5}, we depict the simulation results for the aforementioned user degree distributions. The  \ac{VF} is of duration $\fraLen=200\, \pkLen$ and the rate is $\rate=1.5$~[b/sym]. Consequently, ${\s=\fraLen/\pkLen=200}$ and according to \eqref{eq:phi}, for the selected \ac{SNR} and rate, $\ifr\cong 0.44$ and thus ${\nVp=\lfloor \fraLen/\Vpd \rfloor=225}$.
\end{sloppypar}
\begin{figure}[!t]
	\centering
	\includegraphics[width=\columnwidth]{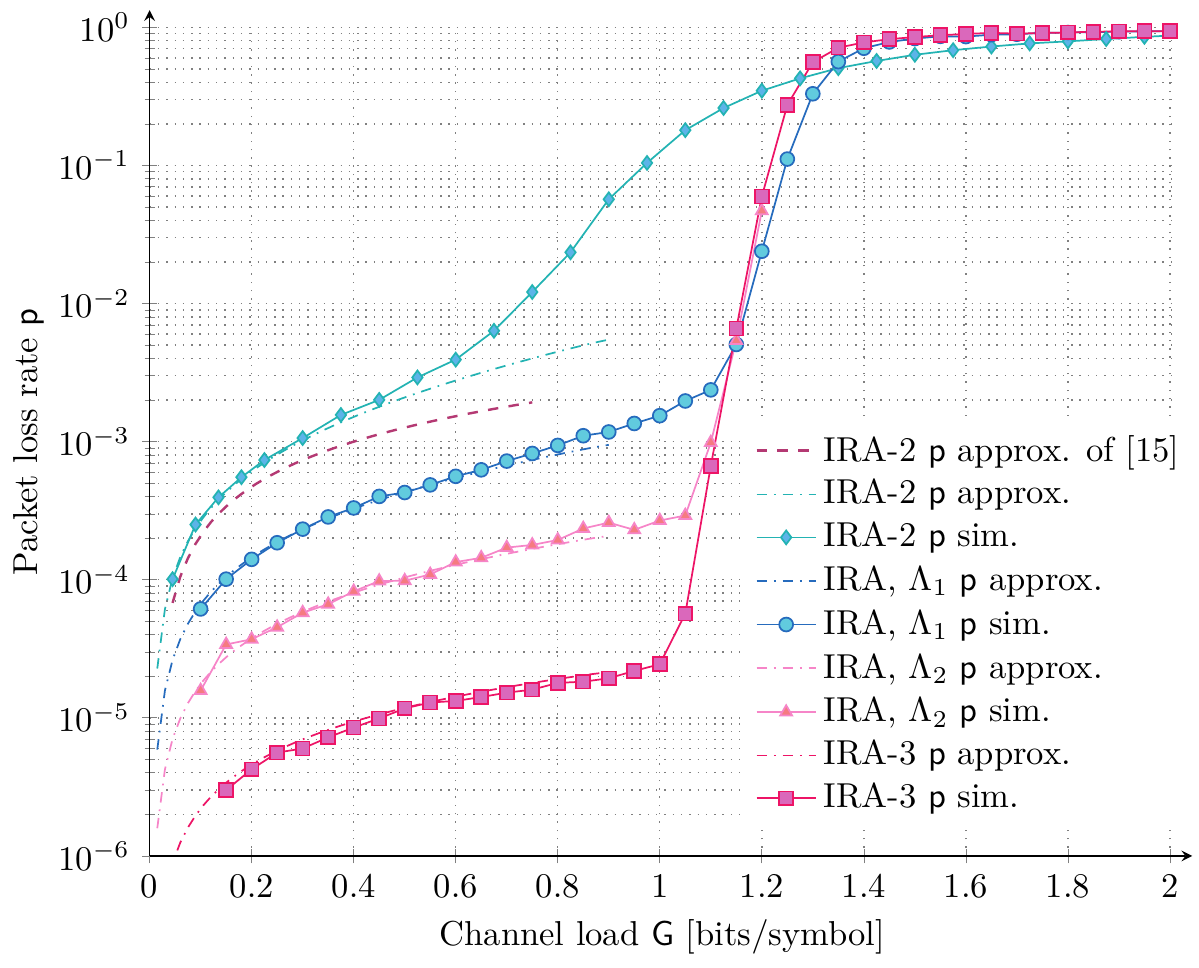}
	\vspace{-4ex}
	\caption{\Ac{PLR} performance simulated (marked solid lines) compared to the derived approximation (dash-dotted lines), for  $\SNR=6$~dB, $\rate=1.5$~[b/sym] and $\fraLen=200\, \pkLen$. Regular user degree distributions with ${\ud=x^2}$ and ${\ud=x^3}$ are denoted with IRA-$2$ and IRA-$3$ respectively. For IRA-$2$ a comparison with the approximation derived in \cite{Clazzer18:ECRA} is also provided (dashed line). The irregular user degree distributions are $\ud_1=0.263x^2+0.344x^3+0.393x^5$ and $\ud_2=0.51x^2+0.49x^4$.}
	\label{fig:n200_R=1.5}
	\vspace{-3ex}
\end{figure}

The solid marked lines correspond to Monte Carlo simulations while dash-dotted lines identify the \ac{PLR} approximation. For reference, the approximation of \cite{Clazzer18:ECRA} corresponding to \eqref{eq:ubCRDSAb} is also depicted for IRA-$2$ (dashed line). Remarkably, the derived \ac{PLR} provides an accurate prediction of the simulated \ac{PLR}  for all  considered user degree distributions in the error floor region. For IRA-2, the derived approximation closes the gap to the simulated curve of  the approximation in \cite{Clazzer18:ECRA}. It is also noteworthy that the approximation of \cite{Clazzer18:ECRA} cannot address any irregular distribution. Finally, we underline that despite the computation of the number of disjoint vulnerable periods $\nVp$ disregards collisions among more than two replicas, the approximation remains particularly tight also in the presence of user degrees larger than two.

In order to investigate the robustness of the error floor approximation, we delve into two slightly modified scenarios. The first one reduces the replicas protection against noise and interference by selecting a larger rate of $\rate=2$~[b/sym] for the same \ac{SNR} of $\SNR=6$~dB and is shown in Fig.~\ref{fig:n200_R=2}. A higher rate results in packets with shorter duration and abbreviates the \ac{VF} duration, but also reduces the interference that can be counteracted. As a result, $\ifr\cong 0.78$ and thus ${\nVp=\lfloor \fraLen/\Vpd \rfloor=127}$. The same color and marker code of Fig.~\ref{fig:n200_R=1.5} is adopted. We can notice that also in this case a very tight match between the derived approximation and the numerical results for all the user degree distributions is obtained. Also the sharp drop in the \ac{PLR} performance at very low channel load is precisely predicted and confirms once more the great potential of the analytical approximation.

\begin{figure}[!t]
	\centering
	\includegraphics[width=\columnwidth]{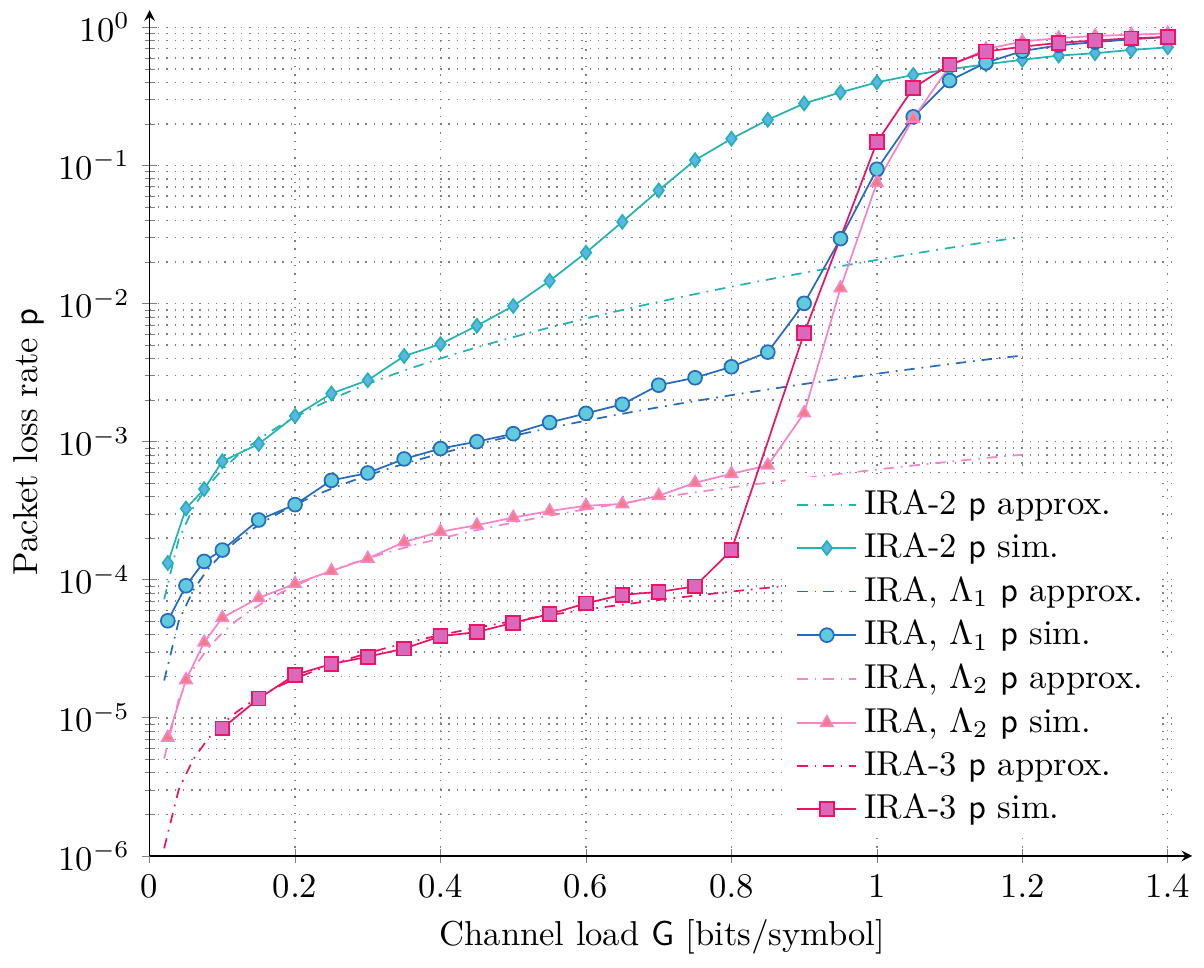}
	\vspace{-4ex}
	\caption{\Ac{PLR} performance simulated (marked solid lines) compared to the derived approximation (dash-dotted lines), for  $\SNR=6$~dB, $\rate=2$~[b/sym] and $\fraLen=200\, \pkLen$. Regular user degree distributions with ${\ud=x^2}$ and ${\ud=x^3}$ are denoted with IRA-$2$ and IRA-$3$ respectively. The irregular user degree distributions are $\ud_1=0.263x^2+0.344x^3+0.393x^5$ and $\ud_2=0.51x^2+0.49x^4$.}
	\label{fig:n200_R=2}
	\vspace{-3ex}
\end{figure}

We show in Fig.~\ref{fig:n100_R=1.5} the second modified scenario which considers a shorter maximum latency among replicas of the same user, i.e., ${\fraLen=100\, \pkLen}$, while keeping the rate ${\rate=1.5}$~[b/sym] and the  $\SNR=6$~dB. Although driven by a different purpose\textemdash reducing the latency among replicas has an important impact on both transmitter and receiver design\textemdash the performance results mirror what we observed when increasing the rate (cf. Fig.~\ref{fig:n200_R=2}). Latency reduction may impact the transmitter architecture by alleviating the battery requirements since the device is required to be active for a shorter period. At the same time, the \ac{VF} duration has a direct influence on the storage capabilities of the receiver. Enabling the \ac{SIC} procedure demands storage availability proportional to the \ac{VF} and thus a reduction eases the memory needs. As a result ${\s=\fraLen/\pkLen=100}$ and $\ifr\cong 0.44$, which provides a number of vulnerable periods of ${\nVp=\lfloor \fraLen/\Vpd \rfloor=112}$.

\begin{figure}[t]
	\centering
	\includegraphics[width=\columnwidth]{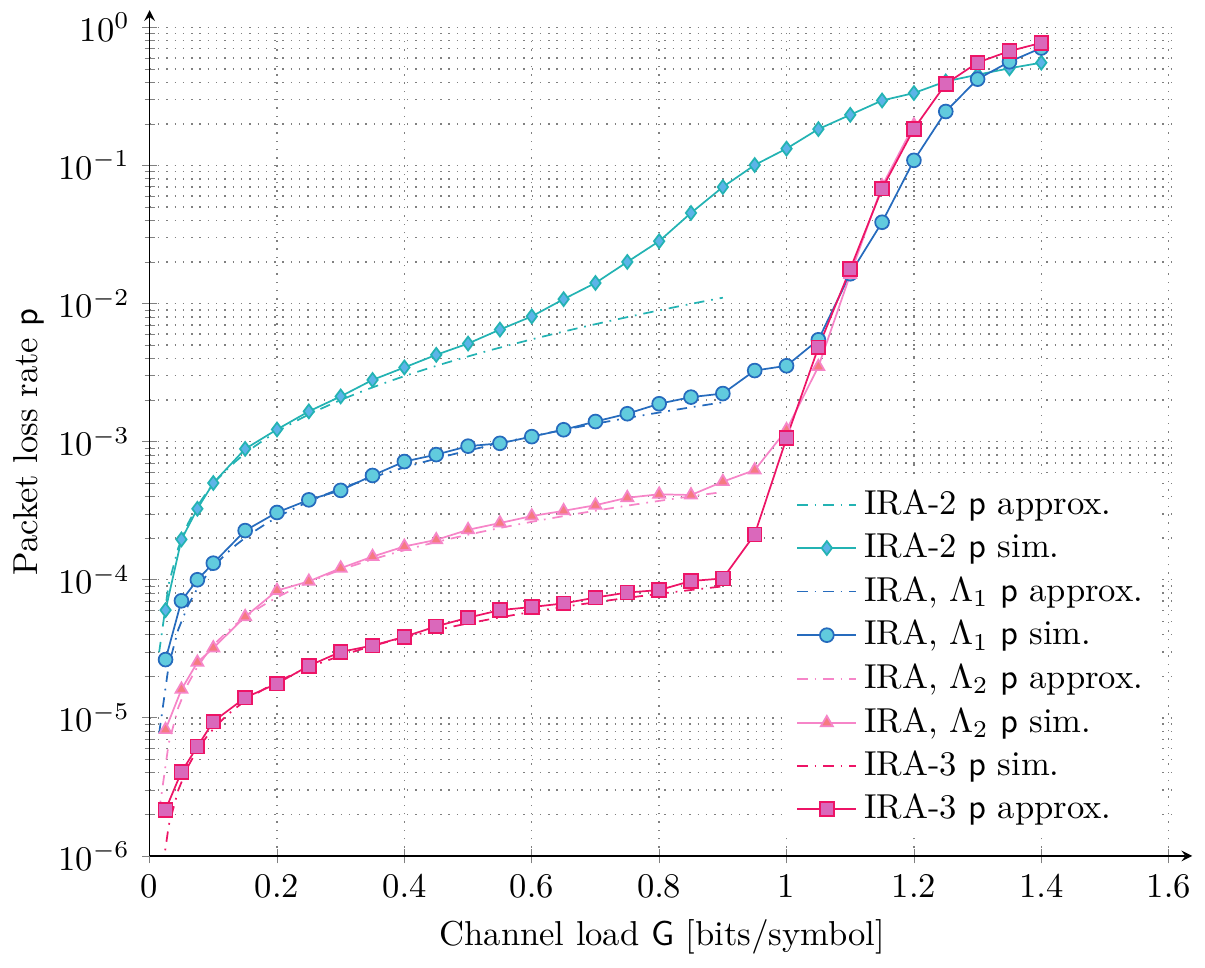}
	\vspace{-5ex}
	\caption{Simulated \Ac{PLR} performance (marked solid lines) and derived approximation (dash-dotted lines), for  $\SNR=6$~dB, $\rate=1.5$~[b/sym] and $\fraLen=100\, \pkLen$. Regular user degree distributions with ${\ud=x^2}$ and ${\ud=x^3}$ are denoted with IRA-$2$ and IRA-$3$ respectively. The irregular user degree distributions are $\ud_1=0.263x^2+0.344x^3+0.393x^5$ and $\ud_2=0.51x^2+0.49x^4$.}
	\label{fig:n100_R=1.5}
	\vspace{-3ex}
\end{figure}

The same color and marker code of Fig.~\ref{fig:n200_R=1.5} are adopted. Also in this case, the error floor is tightly predicted for all the user degree distributions. Compared to the scenario of Fig.~\ref{fig:n200_R=2}, the simulated \ac{PLR} experiences a larger range of channel load values pertaining to the error floor regime. In particular, for IRA-$2$ the range is extended from $\lo=0.4$~[b/sym] to $\lo=0.6$~[b/sym], while for all other distribution is expanded from $\lo=0.75$~[b/sym] to $\lo=0.9$~[b/sym]. A shorter \ac{VF} corresponds to a smaller \ac{MAC} frame in the slot-synchronous equivalent systems  \ac{CRDSA}~\cite{Casini2007} and \ac{IRSA}~\cite{Liva2011}. It is known from the literature that smaller \ac{MAC} frames worsen the \ac{PLR} performance. Similarly, we can also observe in the case of asynchronous \ac{RA} that lowering the \ac{VF} duration has a detrimental impact on the \ac{PLR} (cf. Fig.~\ref{fig:n200_R=1.5} and Fig.~\ref{fig:n100_R=1.5}).\footnote{It shall be noted that ideal interference cancellation has been assumed. When replicas are not perfectly canceled, a residual interference level may be experienced by not yet resolved packets which degrades the performance of the overall scheme.}

\section{Conclusion}
Driven by  \ac{IoT} applications, in this paper we considered an asynchronous \ac{RA} scheme employing irregular packet replication and \ac{SIC} at the receiver. We derived an analytical approximation of the \ac{PLR} of \ac{IRA} that accurately predicts its performance in the error floor region. The accuracy of the \ac{PLR} prediction is demonstrated via Monte Carlo simulations for a number of different scenarios. The possibility to analytically evaluate the \ac{PLR} is particularly appealing, since
error floors in the order of $10^{-5}$ and lower are achievable and numerical evaluations become particularly lengthy. Furthermore, the derived approximation can be used to optimize the degree distribution of \ac{IRA} for a given scenario of interest in terms of \ac{SNR}, rate, \ac{VF} duration, and operative channel load value to achieve a given \ac{PLR}.

\bibliographystyle{IEEEtran}
\bibliography{IEEEabrv,References}

\end{document}

%% file: acronyms2.tex
\DeclareAcronym{AWGN}{short = AWGN ,long = additive white Gaussian noise}
\DeclareAcronym{ACRDA}{short = ACRDA ,long = asynchronous contention resolution diversity ALOHA}
\DeclareAcronym{CDF}{short = CDF ,long = cumulative distribution function}
\DeclareAcronym{CRA-CC}{short = CRA-CC ,long = CRA-convolutional code}
\DeclareAcronym{CRA-SH}{short = CRA-SH ,long = CRA-shannon bound}
\DeclareAcronym{CRA}{short = CRA ,long = contention resolution ALOHA}
\DeclareAcronym{CRDSA}{short = CRDSA ,long = contention resolution diversity slotted ALOHA}
\DeclareAcronym{CRDSA++}{short = CRDSA++ ,long = contention resolution diversity slotted ALOHA++}
\DeclareAcronym{CRI}{short = CRI ,long = contention resolution interval}
\DeclareAcronym{CSA}{short = CSA ,long = coded slotted ALOHA}
\DeclareAcronym{CSI}{short = CSI ,long = channel state information}
\DeclareAcronym{DAMA}{short = DAMA ,long = demand assigned multiple access}
\DeclareAcronym{DSA}{short = DSA ,long = diversity slotted ALOHA}
\DeclareAcronym{DSSS}{short = DSSS ,long = direct sequence spread spectrum}
\DeclareAcronym{E_SSA}{short = E-SSA ,long = enhanced spread spectrum ALOHA}
\DeclareAcronym{ECRA}{short = ECRA ,long = enhanced contention resolution ALOHA}
\DeclareAcronym{ECRA-SC}{short = ECRA-SC ,long = ECRA selection combining}
\DeclareAcronym{ECRA-MRC}{short = ECRA-MRC ,long = ECRA maximal-ratio combining}
\DeclareAcronym{EGC}{short = EGC ,long = equal-gain combining}
\DeclareAcronym{FEC}{short = FEC ,long = forward error correction}
\DeclareAcronym{GEO}{short = GEO ,long = geostationary orbit}
\DeclareAcronym{IC}{short = IC ,long = interference cancellation}
\DeclareAcronym{IoT}{short = IoT ,long = Internet of things}
\DeclareAcronym{IRA}{short = IRA ,long = irregular repetition ALOHA}
\DeclareAcronym{IRCRA}{short = IRCRA ,long = irregular repetition contention resolution ALOHA}
\DeclareAcronym{IRSA}{short = IRSA ,long = irregular repetition slotted ALOHA}
\DeclareAcronym{LDPC}{short = LDPC ,long = low-density parity-check}
\DeclareAcronym{M2M}{short = M2M ,long = machine-to-machine}
\DeclareAcronym{MAC}{short = MAC ,long = medium access}
\DeclareAcronym{MF}{short = MF ,long = matched filter}
\DeclareAcronym{MF-TDMA}{short = MF-TDMA ,long = multi-frequency time division multiple access}
\DeclareAcronym{MRC}{short = MRC ,long = maximal-ratio combining}
\DeclareAcronym{MUD}{short = MUD ,long = multiuser detection}
\DeclareAcronym{PDF}{short = PDF ,long = probability density function}
\DeclareAcronym{PER}{short = PER ,long = packet error rate}
\DeclareAcronym{PLR}{short = PLR ,long = packet loss rate}
\DeclareAcronym{QPSK}{short = QPSK ,long = quadrature phase-shift keying}
\DeclareAcronym{RA}{short = RA ,long = random access}
\DeclareAcronym{RCB}{short = RCB ,long = random coding bound}
\DeclareAcronym{RTT}{short = RTT ,long = round trip time}
\DeclareAcronym{SA}{short = SA , long = slotted ALOHA}
\DeclareAcronym{SB}{short = SB ,long = Shannon bound}
\DeclareAcronym{SC}{short = SC ,long = selection combining}
\DeclareAcronym{SIC}{short = SIC ,long = successive interference cancellation}
\DeclareAcronym{SNIR}{short = SNIR ,long = signal-to-noise and interference ratio}
\DeclareAcronym{SINR}{short = SINR ,long = signal-to-interference and noise ratio}
\DeclareAcronym{SNR}{short = SNR ,long = signal-to-noise ratio}
\DeclareAcronym{TDMA}{short = TDMA ,long = time division multiple access}
\DeclareAcronym{UCP}{short = $\Code$-UCP ,long = $\Code$-unresolvable collision pattern}
\DeclareAcronym{VF}{short = VF ,long = virtual frame}

%% file: notation.tex
\newtheorem{definition}{Definition}

\renewcommand{\Pr}{\ensuremath{\text{Pr}}}

\newcommand{\lo}{\mathsf{G}}
\newcommand{\plr}{\mathsf{p}}
\newcommand{\rate}{\mathsf{R}}

\newcommand{\dg}{\mathsf{d}}
\newcommand{\dgm}{\bar{\dg}}
\newcommand{\dgM}{\mathsf d_m}
\newcommand{\ud}{\Lambda}

\newcommand{\numBit}{k}
\newcommand{\numSym}{n_s}

\newcommand{\fraLen}{T_f}
\newcommand{\pkLen}{T_\text{p}}

\newcommand{\rx}{y}
\newcommand{\rxVec}{\bm{\rx}}
\newcommand{\symVal}{x}
\newcommand{\symVec}{\bm{\symVal}}
\newcommand{\intVal}{z}
\newcommand{\intVec}{\bm{\intVal}}
\newcommand{\noise}{n}
\newcommand{\noiseVec}{\bm{\noise}}
\newcommand{\noiseVar}{\sigma_{\noise}^2}
\newcommand{\sinr}{\gamma}
\newcommand{\sinrVec}{\Gamma}
\newcommand{\intNum}{m}

\newcommand{\mutInf}{\mathsf{I}}
\newcommand{\avMutInf}{\bar{\mutInf}}
\newcommand{\rvDec}{\mathcal{D}}

\newcommand{\UCP}{\mathcal{S}}
\newcommand{\AllUCP}{\mathcal{A}}
\newcommand{\DUCP}{\AllUCP^*}
\newcommand{\Code}{\mathcal{C}}
\newcommand{\CollClus}{\mathcal{U}}

\newcommand{\upf}{\nu}
\newcommand{\up}{\boldsymbol{\upf}}
\newcommand{\profileS}{\up^\UCP}
\newcommand{\spf}{\mu}

\newcommand{\RStart}{\tau}
\newcommand{\VL}{\RStart_l^*}
\newcommand{\VR}{\RStart_r^*}
\newcommand{\Vg}{\RStart^*}
\newcommand{\Vpd}{T_v}
\newcommand{\nVp}{n_v}

\newcommand{\ifr}{\varphi}

\newcommand{\Pw}{\mathsf{P}}
\newcommand{\intPw}{\mathsf{Z}}
\newcommand{\Ns}{\mathsf{N}}

\newcommand{\us}{u}
\newcommand{\replica}{r}
\newcommand{\tus}{m}
\newcommand{\s}{n_p}

\newcommand{\ps}{\Pr\left(\us \in \UCP |\tus\right)}

\newcommand{\as}{a(\tus,\profileS,\ud)}
\newcommand{\bs}{b(\nVp,\spf^\UCP)}
\newcommand{\cs}{c(\UCP)}
\newcommand{\ds}{d(\nVp,\profileS)}

\newcommand{\nus}{\nu(\Sset)}
\newcommand{\mus}{\mu(\Sset)}
\newcommand{\nusl}{\nu_l(\Sset)}

\newcommand{\Lbar}{{\bar L}}
\newcommand{\X}{{\mathcal X}}
\newcommand{\Y}{{\mathcal Y}}
\newcommand{\AX}{\mathcal{A}_{\mathcal X}}
\newcommand{\PX}{\mathcal{P}_{\mathcal X}}

\newcommand{\xtilde}{\tilde{\boldsymbol{x}}}
\newcommand{\dmin}{d_{E,\mathrm{min}}}
\newcommand{\dE}{d_{E}}
\newcommand{\zero}{{\mathbf{0}}}
\newcommand{\dhmin}{d_{\mathrm{min}}}
\newcommand{\dfree}{d_{\mathrm{free}}}
\newcommand{\dtemp}{d_{\mathrm{temp}}}
\newcommand{\dc}{d_{\mathrm{c}}}
\renewcommand{\dh}{d_{\mathrm{H}}}
\newcommand{\uu}{\boldsymbol{u}}
\newcommand{\x}{\boldsymbol{x}}
\newcommand{\w}{\boldsymbol{w}}
\newcommand{\m}{\boldsymbol{m}}
\newcommand{\xhat}{\hat{\boldsymbol{x}}}
\newcommand{\barr}{\bar{\boldsymbol{r}}}
\newcommand{\barrr}{\bar{r}}
\newcommand{\G}{\mathcal{G}}
\renewcommand{\H}{\boldsymbol{H}}
\newcommand{\Htilde}{\tilde{\boldsymbol{H}}}
\newcommand{\PP}{\boldsymbol{P}}
\newcommand{\I}{\boldsymbol{I}}
\newcommand{\g}{\boldsymbol{g}}
\renewcommand{\a}{\boldsymbol{a}}
\renewcommand{\b}{\boldsymbol{b}}
\newcommand{\q}{\boldsymbol{q}}
\newcommand{\p}{\boldsymbol{p}}
\newcommand{\R}{\boldsymbol{R}}
\newcommand{\Cperp}{\mathcal{C}_{\perp}}
\newcommand{\HH}{\mathbf{H}}
\newcommand{\EbNounc}{\left[\frac{E_b}{N_0}\right]_{\mathrm{unc}}}
\newcommand{\EbNocod}{\left[\frac{E_b}{N_0}\right]_{\mathrm{cod}}}

\renewcommand{\c}{\mathbf{c}}

\newcommand{\bb}{{\boldsymbol{b}}}
\newcommand{\y}{\boldsymbol{y}}
\newcommand{\n}{\boldsymbol{n}}
\newcommand{\rr}{\boldsymbol{r}}
\newcommand{\e}{\boldsymbol{e}}
\newcommand{\Ss}{\mathcal{S}}
\newcommand{\T}{^{\mathsf{T}}}
\newcommand{\shat}{\hat{\boldsymbol{s}}}
\newcommand{\sprime}{\boldsymbol{s}{'}}
\newcommand{\Amin}{A_{\mathrm{min}}}
\newcommand{\AAbmin}{\bar{A}^b_{\mathrm{min}}}

\newcommand{\na}{\bar{n}_{\mathsf a}}
\newcommand{\PB}{P_{\mathsf F}}
\newcommand{\PBcrdsa}{P^\mathsf{CRDSA}_{\mathsf F}}
\newcommand{\epss}{\epsilon^*}
\newcommand{\gs}{g^*}
\newcommand{\aepsr}{\alpha_{\epsilon,R}}
\renewcommand{\b}{\mathsf{b}}
\newcommand{\agr}{\mathsf{a}_{g,R}}
\newcommand{\agz}{\alpha_{g,0}}

\newcommand{\C}{\mathcal{C}}
\newcommand{\lC}{\lambda_{\C}}
\newcommand{\rC}{\rho_{\C}}

\newcommand{\Sset}{\mathcal{S}} 

\newcommand{\Astar}{\mathcal{A}^{\star}}

\newcommand{\nv}{n_\mathsf{v}}
\newcommand{\nvone}{n_\mathsf{v1}}
\newcommand{\nvtwo}{n_\mathsf{v2}}
\newcommand{\dv}{d_\mathsf{v}}

\newcommand{\Rf}{R_\mathsf{f}}
\newcommand{\Rione}{R_\mathsf{i1}}
\newcommand{\Ritwo}{R_\mathsf{i2}}
\newcommand{\Rithree}{R_\mathsf{i3}}

\newcommand{\asc}{a_{\mathsf{CRDSA}}(\Sset,m)}
\newcommand{\bsc}{b_{\mathsf{CRDSA}}(\Sset)}
\newcommand{\dsc}{d_{\mathsf{CRDSA}}(\Sset)}

\newcommand{\SNR}{\mathsf{P}/\mathsf{N}}

\newcommand{\Col}{\mathsf{C}}
\newcommand{\bCol}{\bar{\mathsf{C}}}
\newcommand{\uA}{\mathsf{A}}
\newcommand{\uB}{\mathsf{B}}
\newcommand{\uC}{\mathsf{\tilde C}}
\newcommand{\Fl}{\mathsf{F_l}}
\newcommand{\Fr}{\mathsf{F_r}}
\newcommand{\Flt}{\mathsf{F_{l2}}}
\newcommand{\Flth}{\mathsf{F_{l3}}}
\newcommand{\Frt}{\mathsf{F_{r2}}}
\newcommand{\Frth}{\mathsf{F_{r3}}}
\newcommand{\phia}{\phi_a}
\newcommand{\pst}{\bar{p}(\mathcal S_2)}
\newcommand{\psth}{\bar{p}(\mathcal S_3)}